\documentclass[preprint,12pt]{aastex}

\def\ltsima{$\; \buildrel < \over \sim \;$}
\def\lsim{\lower.5ex\hbox{\ltsima}}
\def\gtsima{$\; \buildrel > \over \sim \;$}
\def\gsim{\lower.5ex\hbox{\gtsima}}
\def\rs{$R_\star = 1.003\pm 0.027$~$R_\odot$}
\def\rp{$R_p = 1.222\pm 0.038$~$R_{\rm Jup}$}
\def\ms{$M_\star = 0.980\pm 0.062$~$M_\odot$}

\def\kms{\ifmmode{\rm km\thinspace s^{-1}}\else km\thinspace s$^{-1}$\fi}

\shortauthors{Holman et al.\ 2006}
\shorttitle{Transit Photometry of TrES-2}

\begin{document}

\bibliographystyle{apj}

\title{
The Transit Light Curve (TLC) Project.\\
VI.~Three Transits of the Exoplanet TrES-2
}

\author{
Matthew J.\ Holman\altaffilmark{1},
Joshua N.\ Winn\altaffilmark{2},
David W.\ Latham\altaffilmark{1},\\
Francis T.\ O'Donovan\altaffilmark{3},
David Charbonneau\altaffilmark{1,6},
Guillermo~Torres\altaffilmark{1},
Alessandro~Sozzetti\altaffilmark{1,4},
Jose Fernandez\altaffilmark{1},
Mark E.\ Everett\altaffilmark{5}
}

\altaffiltext{1}{Harvard-Smithsonian Center for Astrophysics, 60
  Garden Street, Cambridge, MA 02138; mholman@cfa.harvard.edu}
\altaffiltext{2}{Department of Physics, and Kavli Institute for
  Astrophysics and Space Research, Massachusetts Institute of
  Technology, Cambridge, MA 02139}
\altaffiltext{3}{California Institute of Technology, 1200 East
  California Blvd., Pasadena, CA 91125}
\altaffiltext{4}{INAF - Osservatorio Astronomico di Torino, 10025 Pino
  Torinese, Italy}
\altaffiltext{5}{Planetary Science Institute, 1700 East Fort Lowell,
  Tucson, AZ 85719}
\altaffiltext{6}{Alfred P.\ Sloan Research Fellow.}

\begin{abstract}

Of the nearby transiting exoplanets that are amenable to detailed
study, TrES-2 is both the most massive and has the largest impact
parameter. 
We present $z$-band photometry of three transits of TrES-2. We
improve upon the estimates of the planetary, stellar, and orbital
parameters, in conjunction with the spectroscopic analysis of the
host star by Sozzetti and co-workers. We find the planetary radius
to be \rp~ and the stellar radius to be \rs. The quoted
uncertainties include the systematic error due to the uncertainty in
the stellar mass (\ms). The timings of the transits have an accuracy
of 25~s and are consistent with a uniform period, thus providing a
baseline for future observations with the NASA {\it Kepler}
satellite, whose field of view will include TrES-2.

\end{abstract}


\section{Introduction}

Careful follow-up observations of nearby transiting planet systems
have revolutionized our understanding of a whole new kind of planet:
hot Jupiters.  They have been used to reveal absorption by atmospheric
atomic sodium ~\citep{Charbonneau.2002} and the presence of an
extended hydrogen exosphere~\citep{Vidal-Madjar.2003} in HD~209458b,
as well as to detect the thermal infrared emission from TrES-1,
HD~209458b, and
HD~189733b~\citep{Charbonneau.2005,Deming.2005a,Deming.2006}.  They
have been used to investigate the spin-orbit alignment of
HD~209458b~\citep{Queloz.2000, Winn.2005} and
HD~189733b~\citep{Winn.2006}.  Most recently, spectra of the infrared
planetary emission of HD~189733b~\citep{Grillmair.2007} and
HD~209458b~\citep{Richardson.2007}, obtained with the {\it Spitzer}
Space Telescope, have been used to constrain models of the atmospheric
content of those planets.

Through these observations we are steadily improving our understanding
of the interior and atmospheric structure of hot Jupiters. Future
measurements, such as reflected-light observations or the detection of
other atmospheric constituents through transmission spectroscopy, will
continue to advance our knowledge of these planets.  One goal of the
Transit Light Curve (TLC) Project is to support these efforts by
refining the estimates of the planetary, stellar, and orbital
parameters, through high-accuracy, high-cadence photometry of
exoplanetary transits. We also seek to measure or bound any variations
in the transit times and light-curve shapes that would be caused by
the influence of additional bodies in the
system~\citep{Miralda-Escude.2002, Agol.2005,Holman.2005}. Along the
way, we are exploring different techniques for photometry and
parameter determination. Previous papers in this series have reported
results for the exoplanets XO-1b~\citep{Holman.2006},
OGLE-TR-111b~\citep{Winn.2007a}, TrES-1~\citep{Winn.2007b},
OGLE-TR-10b~\citep{Holman.2007}, and HD~189733b~\citep{Winn.2007c}.

The present paper is concerned with TrES-2, the second transiting hot
Jupiter discovered by the Trans-atlantic Exoplanet Survey
\citep{O'Donovan.2006b}. The planet orbits a nearby G0~V star
(GSC~03549-02811) and transits every $\sim$2.5~days. Although each of
the fourteen known transiting exoplanets has its own story to tell,
(see \citealt{Charbonneau.2007} for a review), the TrES-2 system has
at least three distinguishing characteristics.

First, TrES-2 is the first transiting extrasolar planet discovered in
the field of view of the NASA {\it Kepler}
mission~\citep{Borucki.2003,Basri.2005}. {\it Kepler} will observe
nearly six hundred transits of TrES-2 during the nominal 4~yr lifetime
of the mission. This opportunity prompts us to improve the
determinations of the orbital parameters of TrES-2 for comparison to
the future estimates from {\it Kepler}.

Second, TrES-2 has the highest impact parameter of any known
nearby transiting extrasolar planet. This makes the duration of the transit
(as well as the duration of ingress and egress) more sensitive to
changes in impact parameter. This, in turn, makes TrES-2 an excellent
target for the detection of long-term changes in transit
characteristics induced by orbital
precession~\citep{Miralda-Escude.2002}.

Third, the mass of TrES-2 is the largest of the known nearby
transiting extrasolar planets. Furthermore, the radius of TrES-2
appears somewhat larger than predicted by simple structural models of
irradiate hot Jupiters, as also appears to be the case for HAT-P-1b,
WASP-1b, and HD~209458b (although see Burrows et al.~2007 for a
contrary view).

In what follows we present TLC results for TrES-2. In \S~2 we describe
the observations and the data reduction procedures. In \S~3 we
describe the model and techniques we used to estimate the physical and
orbital parameters of the TrES-2 system, and in \S~4 we summarize our
results.

\section{The Observations and Data Reduction}

We observed four transits of TrES-2.  According to the ephemeris
provided by \citet{O'Donovan.2006b},
\begin{equation}
T_c(E) = 2,453,957.6358~\mathrm{[HJD]} + E\times (2.47063~\mathrm{days}),
\label{eq:ephemeris}
\end{equation}
these transits correspond to epochs 13, 15, 32, and 34 on
UT~2006~Sept~11, Sept~16, Oct~28, and Nov~2, respectively.
Observations of a fifth transit, epoch 17, were scheduled but were
not executed due to poor weather. 

We observed these transits with KeplerCam on the 1.2m (48~inch)
telescope of the Fred L.\ Whipple Observatory (FLWO) on Mt.~Hopkins,
Arizona.  This camera (P.I.\ D.~Latham) was built for a photometric
survey of the target field of the {\it Kepler}\/ satellite
mission~\citep{Borucki.2003}. It has a single
$4\mathrm{K}\times4\mathrm{K}$ Fairchild 486 CCD with a $23\farcm 1
\times 23\farcm 1$ field of view.  We used $2 \times 2$ binning, for
which the readout and reset time is 11.5~s and the typical read noise
is 7~$e^{-}$ per binned pixel.  The response of each amplifier
deviates from linearity by less that 0.5\% over the range of counts
from the faintest to brightest comparison star.  We observed through
the SDSS $z$ filter, the reddest available band, in order to minimize
the effect of color-dependent atmospheric extinction on the relative
photometry, and to miminize the effect of limb-darkening on the
transit light curve.

The full-width at half-maximum (FWHM) of a stellar image was typically
$\sim$3 binned pixels ($2\arcsec$) on Sept~11, Sept~16, and Nov~2; the
FWHM ranged from $\sim$3 to $\sim$8~pixels on Oct~28.  We used
automatic guiding to maintain the locations of TrES-2 and its
comparison stars to within a few pixels over the course of each night.
We repeatedly took 30~s exposures for 3.5--5~hr bracketing the
predicted transit midpoint. The conditions on UT~2006~Sept~11 were
clear during the time of the observations, and the images were taken
through airmasses ranging from 1.05 to 1.90.  The conditions on
UT~2006~Sept~16 were also clear, and the airmass ranged from 1.05 to
1.40.  There were clouds passing overhead during the observations on
UT~2006~Oct~28, and the airmass ranged from 1.05 to 2.50.  The
observing conditions were significantly worse during and after egress;
the result was essentially observations of only a partial transit. 
Consequently, the data from Oct~28 were not included in the analysis
below.   There were very thin clouds during the observations on
UT~2006~Nov~2, and the airmass ranged from 1.15 to 1.95.

The images were calibrated using standard IRAF\footnote{IRAF is
  distributed by the National Optical Astronomy Observatories, which
  are operated by the Association of Universities for Research in
  Astronomy, Inc., under cooperative agreement with the National
  Science Foundation.} procedures for the overscan correction,
trimming, bias subtraction, and flat-field division.  We did not
attempt to correct the fringing that was apparent with the $z$ filter.
The fringing had a small amplitude and little effect on the final
photometry, given the accuracy of the automatic guiding.  We then
performed aperture photometry of TrES-2 and 20 nearby comparison
stars, using an aperture radius of 8.0 pixels ($4\farcs 3$) for each
night. We subtracted the underlying contribution from the sky, after
estimating its brightness within an annulus ranging from 30 to 35
pixels in radius, centered on each star.  We divided the flux of
TrES-2 by the total flux of the comparison stars.

To estimate the uncertainties in our photometry, we computed the
quadrature sum of the errors due to Poisson noise of the stars (both
TrES-2 and the comparison stars), Poisson noise of the sky background,
readout noise, and scintillation noise (as estimated according to the
empirical formulas of~\citealt{Young.1967} and
\citealt{Dravins.1998}). The dominant term is the Poisson noise from
TrES-2. The final time series is plotted in Fig.~1 and is available in
electronic form in Table~1. (In that table, the quoted errors have
been rescaled such that $\chi^2/N_{\rm dof} = 1$ for the best-fitting
model, as explained in the next section.)

\begin{figure}[p]
\epsscale{0.85}
\plotone{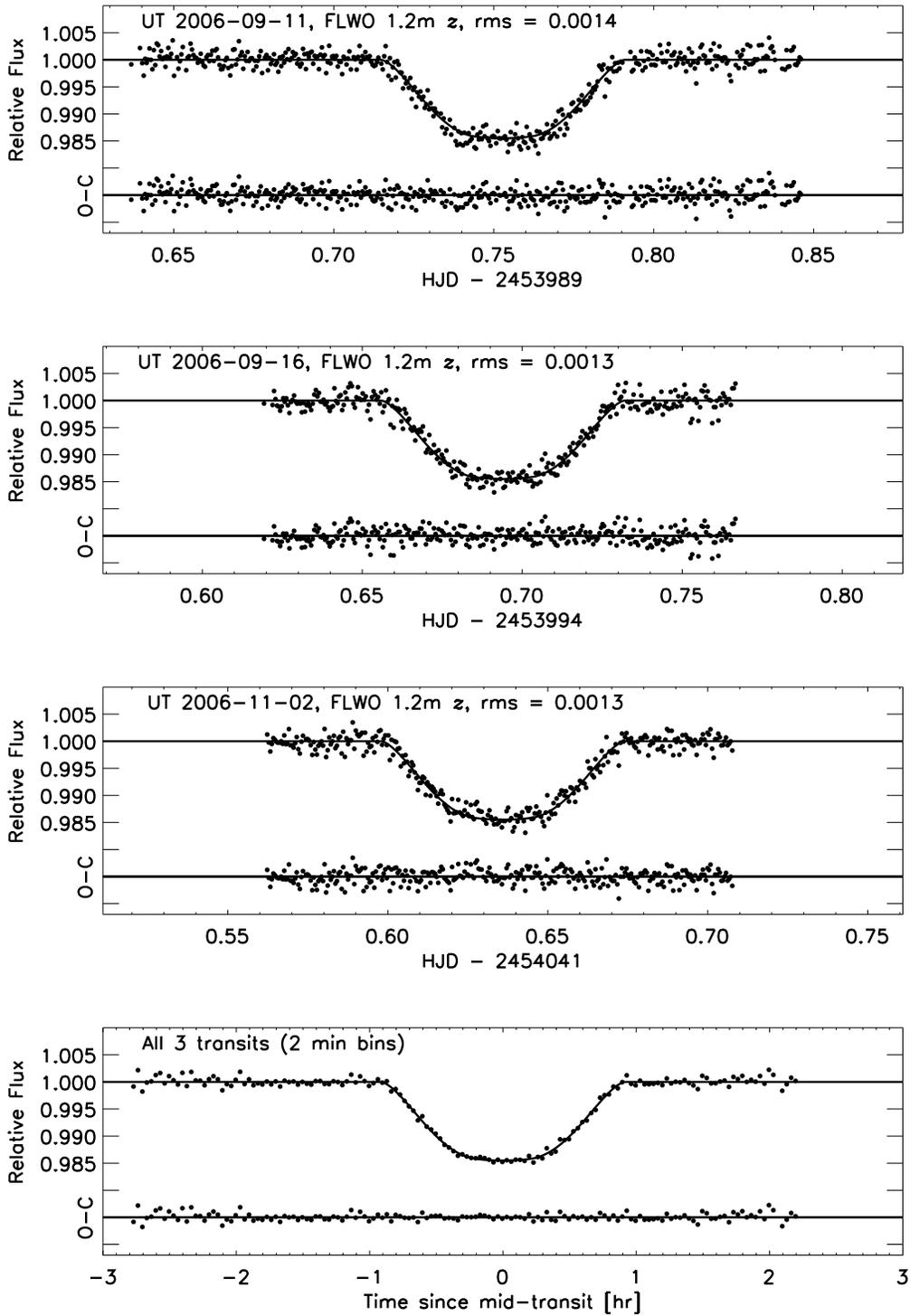}
\caption{
Photometry of TrES-2 in the $z$ band, using
the FLWO 1.2m telescope and Keplercam.  These data were used to
estimate the planetary, stellar, and orbital parameters (see \S~3).
The bottom panel is a composite light curve created from the three data
sets, after time-shifting and averaging into 2~min bins.  The
residuals (observed$-$calculated) are plotted beneath the data. 
\label{fig:1}}
\end{figure}

\section{Determination of System Parameters}

Our methodology for determining the system parameters has been
described in previous TLC
papers~\citep{Holman.2006,Winn.2007a,Winn.2007b,Holman.2007,Winn.2007c},
and is summarized here. We assume a circular orbit of a planet (mass
$M_p$, radius $R_p$) and a star ($M_\star$, $R_\star$), with period
$P$ and inclination $i$ relative to the sky plane. We allow each
transit to have an independent value of $T_c$ (the transit midpoint)
rather than forcing them to be separated by exact multiples of the
orbital period. Thus, the only effect of $P$ on the model is to
determine the semimajor axis $a$ for a given value of ($M_\star +
M_p$). We fixed $P=2.47063$~days \citep{O'Donovan.2006b}; the
uncertainty of $0.00001$~days was negligible for this purpose.

To calculate the relative flux as a function of the projected
separation of the planet and the star, we employed the analytic
formulas of Mandel \& Agol~(2002), using a quadratic limb darkening
law,
\begin{equation}
\frac{I_\mu}{I_1} = 1 - u_1(1-\mu) - u_2(1-\mu)^2,
\end{equation}
where $I$ is the intensity and $\mu$ is the cosine of the angle
between the line of sight and the normal to the stellar surface. We
chose the values $u_1=0.22$, $u_2=0.32$, based on the tabulated values
of \citet{Claret.2004} and the estimates by \citet{Sozzetti.2007} of
the stellar effective temperature, surface gravity, and metallicity.
We accounted for the color-dependent {\it residual} airmass effects
with a parameter $k$ specific to each transit, such that the observed
flux is equal to the intrinsic (zero airmass) flux times $\exp(-kz)$,
where $z$ is the airmass.  The best-fitting values of $k$ were 0.0021,
0.0086, and -0.0005, for Sept~11, Sept~16, and Nov~2, respectively. We
also fitted for the out-of-transit flux $f_{\rm oot}$. 

The light curves cannot be used to determine both the stellar mass and
radius; there is a fitting degeneracy $R_\star \propto
M_\star^{1/3}$. Our usual approach is to assume a value for $M_\star$
(based on external analyses of the stellar spectrum) and then
determine $R_\star$ by fitting the light curves. This case was
slightly different because we worked in conjunction with
\citet{Sozzetti.2007}, who sought to {\it improve}\, the estimates of
the stellar parameters based on the results of the light-curve fit. We
worked iteratively, as described below in more detail; for our final
analysis, we fixed $M_\star=0.98~M_\odot$.

Our fitting statistic was \begin{equation}
\chi^2 =
\sum_{j=1}^{N_f}
\left[
\frac{f_j({\mathrm{obs}}) - f_j({\mathrm{calc}})}{\sigma_j}
\right]^2
,
\label{eq:chi2}
\end{equation}
where $f_j$(obs) is the flux observed at time $j$, $\sigma_j$ controls
the relative weights of the data points, and $f_j$(calc) is the
calculated value. It is important for $\sigma_j$ to include
measurement errors and also any unmodeled systematic effects, and in
particular to account for time-correlated noise, which effectively
reduces the number of independent measurements. Our approach was as
follows. First, we rescaled the instrumental uncertainties such that
$\chi^2/N_{\rm dof} = 1$ for the best-fitting model. Table~1 lists the
resulting uncertainties. Second, we followed the procedure of
\citet{Gillon.2006} to decompose the observed noise into ``white
noise'' (that which averages down as $1/\sqrt{N}$, where $N$ is the
number of data points) and ``red noise'' (that which does not average
down over some specified time interval). Specifically, we calculated
the standard deviation of the residuals ($\sigma$) and the standard
deviation of the time-averaged residuals ($\sigma_N$). The averaging
time was 1~hr (a timescale comparable to the transit event),
corresponding to a number $N$ of data points that depended upon the
cadence of observations.  Then we solved for the white noise
$\sigma_w$ and red noise $\sigma_r$ from the system of equations
\begin{eqnarray}
\sigma_1^2 & = & \sigma_w^2 + \sigma_r^2, \\
\sigma_N^2 & = & \frac{\sigma_w^2}{N} + \sigma_r^2.
\end{eqnarray}
Finally, to account approximately for the effective reduction in the
number of independent data points, we rescaled the $\sigma_j$ in
Eq.~(\ref{eq:chi2}) by the factor $\sigma_r/(\sigma_w/\sqrt{N})$. In
this case, the Sept~11 and Nov~2 transits did not show evidence for
red noise according to this criterion, but for the Sept~16 transit the
red-noise rescaling factor was 1.14.  For that transit, we find
$sigma_r = 0.00016$ and $\sigma_w = 0.0014$.  To be conservative, we a
pplied this same factor 1.14 to the data from all 3 transits.

In short there were 12 model parameters: $\{R_\star, R_p, i\}$, as
well as $\{T_c, k, f_{\rm oot}\}$ for each of 3 transits.  We
determined the {\it a posteriori} probability distributions for these
parameters using the same Markov Chain Monte Carlo algorithm described
in previous TLC papers.  We took the median value of each probability
distribution to be the best estimate of each parameter, and the
standard deviation to be the 1~$\sigma$ uncertainty. In addition to
this statistical error, for the special cases of $R_\star$ and $R_p$
there is an additional error arising from the uncertainty in
$M_\star$, which we add to the statistical error in quadrature.

Our choice of $M_\star$ merits further discussion since it is based on
a novel iterative procedure conducted in tandem with
\citet{Sozzetti.2007}. The underlying idea is that when fitting a
light curve, the results for $R_\star$ and $R_p$ depend on the choice
of $M_\star$, while the result for $R_\star/a$ is {\it independent}\,
of $M_\star$ because both $R_\star$ and $a$ vary as $M_\star^{1/3}$
for a fixed value of the orbital period. (There is, however, a minor
dependence of $R_\star/a$ on the choice of limb darkening function,
which is in turn informed by the estimates of the stellar parameters.)
Meanwhile, as \citet{Sozzetti.2007} have shown, $R_\star/a$ is useful
for estimating $M_\star$, since it can be directly related to the
stellar mean density through Kepler's Third Law (see also
\citet{Seager.2003}):
\begin{equation}
\frac{a}{R_\star} =
\left(\frac{G P^2}{4\pi^2}\right)^{1/3}
\left(\frac{M_\star + M_p}{R_\star^3}\right)^{1/3}.
\end{equation}
This makes $R_\star/a$ a useful proxy for $\log g$ for the purpose of
comparing the observed stellar properties with theoretical isochrones.
The advantage of $R_\star/a$ is that in typical cases it is more
precisely determined than the spectroscopic value of $\log g$.

We iterated as follows. First, we fitted the light curves using the
choices $M_\star=1.08~M_\odot$, $u_1=0.18$, $u_2=0.34$, based on the
previous estimates of the relevant stellar parameters by
\citet{O'Donovan.2006b}. Next, we passed our results for $R_\star/a$
to \citet{Sozzetti.2007}, who used it to refine the estimate of
$M_\star$. (We refer the reader to \citet{Sozzetti.2007} for details
on how this refinement was achieved.)  In return,
\citet{Sozzetti.2007} provided us with a new estimate of $M_\star$,
along with a new estimate of the stellar surface gravity (which
affects the choice of limb darkening law).  We refitted the light
curves using the updated values of the stellar mass and the slightly
adjusted limb darkening law. Then we passed our new result for
$R_\star/a$ back to \citet{Sozzetti.2007}, who used it to refine the
estimate of $M_\star$ and $\log g$, and so forth. This process
converged after a few iterations, leading to the final choices for
$M_\star$, $u_1$, and $u_2$ noted above.

While it is possible for the value of the stellar radius that
minimizes $\chi^2$ to be inconsistent with the theoretical mass-radius
relation, in this case we have effectively required consistency with
the theoretical mass-radius relation by iterating with
\citet{Sozzetti.2007}.

\section{Results}

The final results are given in Table~\ref{tbl:params}. In addition to
the results for the basic model parameters, we have also included in
this table a number of interesting derived quantities, such as
$a/R_\star$ (which is related to the stellar mean density, as
described above) and the calculated durations of the transit and the
partial transit phases. The most interesting parameters are the radius
of the star, the radius of the planet, the orbital inclination, and
the mid-transit times, which we discuss in turn.

We find the stellar radius to be \rs, where the quoted error includes
both the statistical error ($0.017$) and the systematic error due to
the uncertainty in the stellar mass ($0.021$).  This estimate agrees
with all of the star's observed broadband colors and spectral
properties as determined by \citet{Sozzetti.2007}, as it must, given
that our analyses were coupled as described in the previous section.

We find the planetary radius to be \rp, where (again) the quoted error
includes both the statistical error ($0.028$) and the systematic
error due to the uncertainty in the stellar mass ($0.026$). The
difference between our value, and the value $R_p =
1.220^{+0.045}_{-0.042}$~$R_{\rm Jup}$ presented by
\citet{Sozzetti.2007}, is slight indeed, although our figure has a
somewhat smaller error bar. The reason why there is any difference at
all is subtle. \citet{Sozzetti.2007} determined $R_p$ by taking our
result for $(R_p/R_\star)$ and the associated uncertainty, and
multiplying by their estimate for $R_\star$ (which in turn was based
on matching the observed values of $T_{\rm eff}$, $a/R_\star$, and
metallicity to theoretical isochrones). In contrast, we determined
$R_p$ and $R_\star$ simultaneously by fitting a parameterized model to
the light curves, as described above, and then accounting for the
uncertainty in the stellar mass. Our analysis takes into account the
correlations between all of the parameters, while that of
\citet{Sozzetti.2007} assumes $(R_p/R_\star)$ is independent of
$a/R_\star$. In this case, our procedure has yielded somewhat more
precise results for $R_p$ and $R_\star$.

For an eclipsing single-lined spectroscopic binary the surface gravity
of the secondary ($GM_p/R_p^2$, in this case) can be determined nearly
independently of any assumptions regarding the properties of the
primary\citep{Southworth.2004,Winn.2007a,Beatty.2007,Sozzetti.2007,Southworth.2007}.
This result holds because the fitting degeneracy for the
radial-velocity data is $M_p \propto M_\star^{2/3}$, and the fitting
degeneracy for the photometric data is $R_p \propto M_\star^{1/3}$,
and in the ratio $M_p/R_p^2$ the stellar mass cancels out. There
remains only a weak dependency of $R_p$ on the choice of limb
darkening law, which is based on knowledge of the host star.  In this
case, the result is $GM_p/R_p^2 = 1976\pm 91$~cm~s$^{-2}$, or $\log
g_p = 3.299\pm 0.020$.

We confirm the finding by \citet{O'Donovan.2006b} that the transit
chord occurs at an unusually large impact parameter, $b \equiv a\cos
i/R_\star = 0.8540 \pm 0.0062$. This is of interest because the error
in the impact parameter is much smaller when the impact parameter is
high than when the transit is near-equatorial (all other things being
equal). This facilitates the detection of small changes in the impact
parameter due to orbital precession, which can be caused by additional
bodies in the system or by the stellar quadrupole field
\citep{Miralda-Escude.2002}. A large impact parameter is also
advantageous for interpreting the Rossiter-McLaughlin effect, as long
as an accurate external measurement of the projected rotation speed of
the star ($v\sin i$) is available \citep{Gaudi.2007}.

Accurate timing of exoplanetary transits is a promising method to
identify additional planets or moons (see, e.g., Holman \& Murray~2005
and Agol et al.~2005), and in this case, transit timing takes on
special importance because TrES-2 is in the field of view of the {\it
  Kepler} mission~\citep{Borucki.2003,Basri.2005}. We have tested
whether or not our 3 measured transit times and the single transit
time reported by \citet{O'Donovan.2006b} are consistent with a uniform
period, by fitting a linear function of epoch number to the observed
times. The residuals to this linear fit are shown in Fig.~2, and are
consistent with zero within the measurement errors. Thus there is not
yet any indication of timing anomalies. Based on our fit, we have
refined the ephemeris. The new value of $T_c$ is
$2,453,957.63479(38)$~[HJD] and the new value of the orbital period is
$2.470621(17)$~days, where the numbers in parentheses are the
1~$\sigma$ uncertainties in the last 2 digits of each figure.

\begin{figure}[p]
\epsscale{1.0} 
\plotone{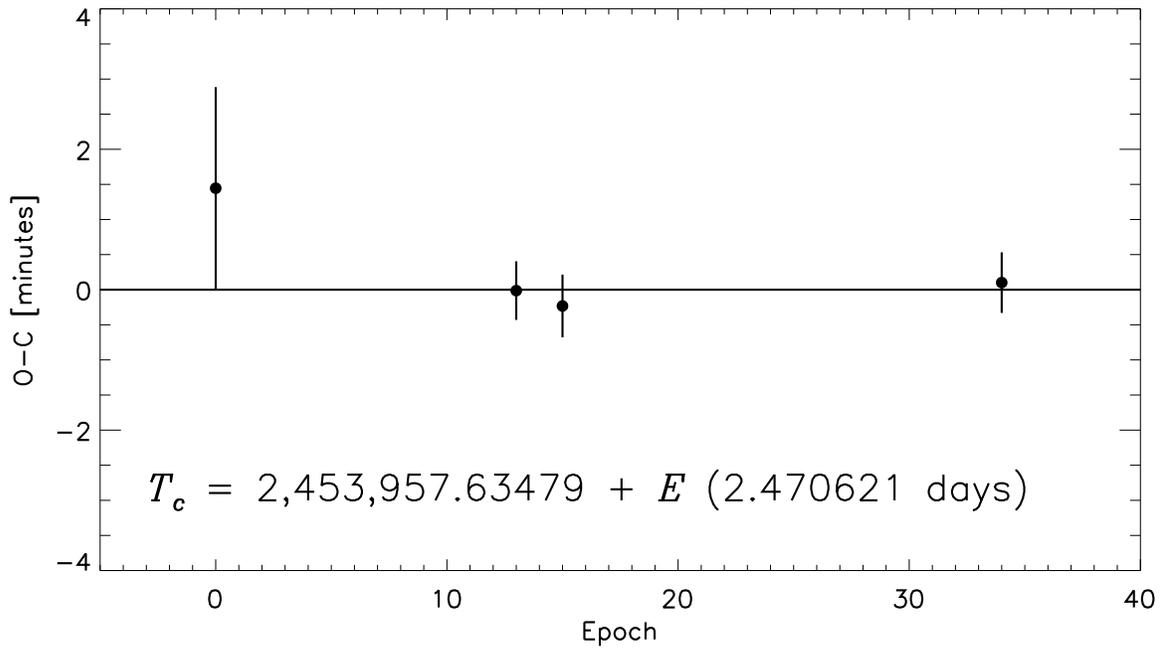}
\caption{
The timing residuals (observed~$-$~calculated) for 4 observed
transits, according to the ephemeris of Eq.~(\ref{eq:ephemeris}).  The
first point corresponds to the $T_c$ reported by
\citet{O'Donovan.2006b}, and the other three points correspond to the
three transits reported in this paper.  The points lie on a horizontal
line, and therefore the data are consistent with a constant period.
\label{fig:tc}}
\end{figure} 

We conducted two tests to check the robustness of our results. First,
we fitted each of the three transits separately, and examined the
scatter in the results. For each of the parameters $\{R_\star, R_p,
i\}$, the 3 different results were all within the 1~$\sigma$
uncertainty of the result when fitting all the transits together. Thus
the results of the 3 transits agree well with one another. Second, we
examined the sensitivity of the results to the limb darkening
function, finding also that the results are robust.  For example, the
effect on $R_p$ of allowing the quadratic limb-darkening coefficients
to be free parameters (rather than fixing them at the values tabulated
by \citet{Claret.2004}) is an increase of 1.0\%. If we use a linear
law instead of a quadratic law, $R_p$ is increased by 0.6\%, and if we
use the four-parameter ``nonlinear'' law of \citet{Claret.2004} (with
coefficients fixed at ATLAS-based values) then $R_p$ is decreased by
0.5\%. None of these changes are very significant compared to the
2.3\% statistical error.

\section{Summary}

Through observations and analysis of three transits, we have improved
upon the estimates of the orbital and physical parameters of TrES-2.
Our results are consistent with the estimates of the stellar and
planetary radii by \citet{O'Donovan.2006b}, but have smaller
uncertainties.  We also show that the available transit times are
consistent with a uniform period.  In our analysis of the photometry
we have made use of an improved estimate of the stellar mass from
\citet{Sozzetti.2007}.  This estimate was obtained by iteratively
combining values of $a/R_\star$ determined from the light curves with
values of effective temperature and metallicity determined from
stellar spectra.  This novel technique can be applied to all
transiting systems for which high quality stellar spectra and high
precision light curves are available.  Our observations and analysis
help lay the ground work for interpreting the $\sim 600$ transits of
TrES-2 that will be observed by {\it Kepler}.

\acknowledgments We thank E.\ Falco for accommodating our observing
schedule changes. KeplerCam was developed with partial support from
the Kepler Mission under NASA Cooperative Agreement NCC2-1390 (P.I.\
D.~Latham), and the KeplerCam observations described in this paper
were partly supported by grants from the Kepler Mission to SAO and
PSI\@.  MJH acknowledges support for this work NASA Origins grant
NG06GH69G. Work by F.T.O'D. and D.C.\ was supported by NASA under
grant NNG05GJ29G, issued through the Origins of Solar Systems Program.
AS gratefully acknowledges the Kepler Mission for partial support under
NASA Cooperative Agreement NCC 2-1390. GT acknowledges partial support
for this work from NASA Origins grant NNG04LG89G.


\begin{thebibliography}{30}
\expandafter\ifx\csname natexlab\endcsname\relax\def\natexlab#1{#1}\fi

\bibitem[{{Agol} {et~al.}(2005){Agol}, {Steffen}, {Sari}, \&
  {Clarkson}}]{Agol.2005}
{Agol}, E., {Steffen}, J., {Sari}, R., \& {Clarkson}, W. 2005, \mnras, 359, 567

\bibitem[{{Basri} {et~al.}(2005){Basri}, {Borucki}, \& {Koch}}]{Basri.2005}
{Basri}, G., {Borucki}, W.~J., \& {Koch}, D. 2005, New Astronomy Review, 49,
  478

\bibitem[{{Beatty} {et al.}(2007)}]{Beatty.2007}
{Beatty}, T.~G. {et al.} 2007, \apj, in press [arXiv:0704.0059]

\bibitem[{{Borucki} {et~al.}(2003){Borucki}, {Koch}, {Basri}, {Caldwell},
  {Caldwell}, {Cochran}, {Devore}, {Dunham}, {Geary}, {Gilliland}, {Gould},
  {Jenkins}, {Kondo}, {Latham}, \& {Lissauer}}]{Borucki.2003}
{Borucki}, W.~J., et al. 
2003, in ASP Conf. Ser. 294: Scientific Frontiers
in Research on Extrasolar Planets, ed. D.~{Deming} \& S.~{Seager}, 427--440

\bibitem[{{Charbonneau} {et~al.}(2005){Charbonneau}, {Allen}, {Megeath},
  {Torres}, {Alonso}, {Brown}, {Gilliland}, {Latham}, {Mandushev}, {O'Donovan},
  \& {Sozzetti}}]{Charbonneau.2005}
{Charbonneau}, D., et al.
2005, \apj, 626, 523

\bibitem[{{Charbonneau} {et~al.}(2007){Charbonneau}, {Brown}, {Burrows}, \&
  {Laughlin}}]{Charbonneau.2007}
{Charbonneau}, D., {Brown}, T.~M., {Burrows}, A., \& {Laughlin}, G. 2007, in
  Protostars and Planets V, ed. B.~{Reipurth}, D.~{Jewitt}, \& K.~{Keil},
  701--716

\bibitem[{{Charbonneau} {et~al.}(2002){Charbonneau}, {Brown}, {Noyes}, \&
  {Gilliland}}]{Charbonneau.2002}
{Charbonneau}, D., {Brown}, T.~M., {Noyes}, R.~W., \& {Gilliland}, R.~L. 2002,
  \apj, 568, 377

\bibitem[{{Claret}(2004)}]{Claret.2004}
{Claret}, A. 2004, \aap, 428, 1001


\bibitem[{{Deming} {et~al.}(2006){Deming}, {Harrington}, {Seager}, \&
  {Richardson}}]{Deming.2006}
{Deming}, D., {Harrington}, J., {Seager}, S., \& {Richardson}, L.~J. 2006,
  \apj, 644, 560

\bibitem[{{Deming} {et~al.}(2005){Deming}, {Seager}, {Richardson}, \&
  {Harrington}}]{Deming.2005a}
{Deming}, D., {Seager}, S., {Richardson}, L.~J., \& {Harrington}, J. 2005,
  \nat, 434, 740

\bibitem[{{Dravins} {et~al.}(1998){Dravins}, {Lindegren}, {Mezey}, \&
  {Young}}]{Dravins.1998}
{Dravins}, D., {Lindegren}, L., {Mezey}, E., \& {Young}, A.~T. 1998, \pasp,
  110, 610

\bibitem[{{Gaudi} \& {Winn}(2007)}]{Gaudi.2007}
{Gaudi}, B.~S. \& {Winn}, J.~N. 2007, \apj, 655, 550

\bibitem[{{Gillon} {et~al.}(2006){Gillon}, {Pont}, {Moutou}, {Bouchy},
  {Courbin}, {Sohy}, \& {Magain}}]{Gillon.2006}
{Gillon}, M., {Pont}, F., {Moutou}, C., {Bouchy}, F., {Courbin}, F., {Sohy},
  S., \& {Magain}, P. 2006, \aap, 459, 249

\bibitem[{{Grillmair} {et~al.}(2007){Grillmair}, {Charbonneau}, {Burrows},
  {Armus}, {Stauffer}, {Meadows}, {Van Cleve}, \& {Levine}}]{Grillmair.2007}
{Grillmair}, C.~J., {Charbonneau}, D., {Burrows}, A., {Armus}, L., {Stauffer},
  J., {Meadows}, V., {Van Cleve}, J., \& {Levine}, D. 2007, \apjl, 658, L115

\bibitem[{{Holman} \& {Murray}(2005)}]{Holman.2005}
{Holman}, M.~J. \& {Murray}, N.~W. 2005, Science, 307, 1288

\bibitem[{{Holman} {et~al.}(2007){Holman}, {Winn}, {Fuentes}, {Hartman},
  {Stanek}, {Torres}, {Sasselov}, {Gaudi}, {Jones}, \&
  {Fraser}}]{Holman.2007}
{Holman}, M.~J., et al.
2007, \apj, 655, 1103

\bibitem[{{Holman} {et~al.}(2006){Holman}, {Winn}, {Latham}, {O'Donovan},
  {Charbonneau}, {Bakos}, {Esquerdo}, {Hergenrother}, {Everett}, \&
  {P{\'a}l}}]{Holman.2006}
{Holman}, M.~J., et al.
2006, \apj, 652, 1715

\bibitem[{{Miralda-Escud{\' e}}(2002)}]{Miralda-Escude.2002}
{Miralda-Escud{\' e}}, J. 2002, \apj, 564, 1019

\bibitem[{{O'Donovan} {et~al.}(2006){O'Donovan}, {Charbonneau}, {Mandushev},
  {Dunham}, {Latham}, {Torres}, {Sozzetti}, {Brown}, {Trauger}, {Belmonte},
  {Rabus}, {Almenara}, {Alonso}, {Deeg}, {Esquerdo}, {Falco}, {Hillenbrand},
  {Roussanova}, {Stefanik}, \& {Winn}}]{O'Donovan.2006b}
{O'Donovan}, F.~T., et al.
2006, \apjl, 651, L61

\bibitem[{{Queloz} {et~al.}(2000){Queloz}, {Eggenberger}, {Mayor}, {Perrier},
  {Beuzit}, {Naef}, {Sivan}, \& {Udry}}]{Queloz.2000}
{Queloz}, D., {Eggenberger}, A., {Mayor}, M., {Perrier}, C., {Beuzit}, J.~L.,
  {Naef}, D., {Sivan}, J.~P., \& {Udry}, S. 2000, \aap, 359, L13

\bibitem[{{Richardson} {et~al.}(2007){Richardson}, {Deming}, {Horning},
  {Seager}, \& {Harrington}}]{Richardson.2007}
{Richardson}, L.~J., {Deming}, D., {Horning}, K., {Seager}, S., \&
  {Harrington}, J. 2007, \nat, 445, 892

\bibitem[{{Seager} \& {Mall{\'e}n-Ornelas}(2003)}]{Seager.2003}
{Seager}, S. \& {Mall{\'e}n-Ornelas}, G. 2003, \apj, 585, 1038

\bibitem[{{Southworth} {et~al.}(2004){Southworth}, {Zucker}, {Maxted},
  \& Smalley}]{Southworth.2004}
{Southworth}, J., {Zucker}, S., {Maxted}, P.~F.~L., \& {Smalley}, B.\
  2004, \mnras, 355, 986

\bibitem[{Southworth} {et~al.}(2007){Southworth}, {Wheatley}, \&
  {Sams}] {Southworth.2007}
{Southworth}, J., {Wheatley}, P., {Sams}, G.\ 2007, \mnras, in press
[arXiv:0704.1570]

\bibitem[{{Sozzetti} {et~al.}(2007){Sozzetti}, {Torres}, {Latham}, {Holman},
  {Winn}, J.B., \& {O'Donovan}~F.T.{Charbonneau}}]{Sozzetti.2007}
{Sozzetti}, A., {Torres}, G., {Charbonneau}, D., {Latham}, D.~W., {Holman}, M.~J., {Winn}, J.~N.,
  Laird, J.B., \& {O'Donovan},~F.T. 2007, \apj, in press

\bibitem[{{Vidal-Madjar} {et~al.}(2003){Vidal-Madjar}, {Lecavelier des Etangs},
  {D{\'e}sert}, {Ballester}, {Ferlet}, {H{\'e}brard}, \&
  {Mayor}}]{Vidal-Madjar.2003}
{Vidal-Madjar}, A., {Lecavelier des Etangs}, A., {D{\'e}sert}, J.-M.,
  {Ballester}, G.~E., {Ferlet}, R., {H{\'e}brard}, G., \& {Mayor}, M. 2003,
  \nat, 422, 143

\bibitem[{{Winn} {et~al.}(2007{\natexlab{a}}){Winn}, {Holman}, \&
  {Fuentes}}]{Winn.2007a}
{Winn}, J.~N., {Holman}, M.~J., \& {Fuentes}, C.~I. 2007{\natexlab{a}}, \aj,
  133, 11

\bibitem[{{Winn} {et~al.}(2007{\natexlab{c}}){Winn}, {Holman}, {Henry},
  {Roussanova}, {Enya}, {Yoshii}, {Shporer}, {Mazeh}, {Johnson}, {Narita}, \&
  {Suto}}]{Winn.2007c}
{Winn}, J.~N., et al.
2007{\natexlab{c}}, \aj, 133, 1828

\bibitem[{{Winn} {et~al.}(2007{\natexlab{b}}){Winn}, {Holman}, \&
  {Roussanova}}]{Winn.2007b}
{Winn}, J.~N., {Holman}, M.~J., \& {Roussanova}, A. 2007{\natexlab{b}}, \apj,
  657, 1098

\bibitem[{{Winn} {et~al.}(2006){Winn}, {Johnson}, {Marcy},
  {Butler}, {Vogt}, {Henry}, {Roussanova}, {Holman}, {Enya}, {Narita}, {Suto},
  \& {Turner}}]{Winn.2006}
{Winn}, J.~N., et al.
2006, \apjl, 653,  L69

\bibitem[{{Winn} {et~al.}(2005){Winn}, {Noyes}, {Holman}, {Charbonneau},
  {Ohta}, {Taruya}, {Suto}, {Narita}, {Turner}, {Johnson}, {Marcy}, {Butler},
  \& {Vogt}}]{Winn.2005}
{Winn}, J.~N., et al.
2005, \apj, 631, 1215



\bibitem[{{Young}(1967)}]{Young.1967}
{Young}, A.~T. 1967, \aj, 72, 747

\end{thebibliography}

\begin{deluxetable}{lcc}
\tabletypesize{\normalsize}
\tablecaption{Relative Photometry of TrES-2\label{tbl:photometry}}
\tablewidth{0pt}

\tablehead{
\colhead{HJD} & \colhead{Relative flux} & \colhead{Uncertainty} \\
}

\startdata
  2453989.63669 &          0.9992 &          0.0013 \\
  2453989.63943 &          1.0022 &          0.0013 \\
  2453989.64013 &          0.9994 &          0.0013 \\
  2453989.64058 &          0.9971 &          0.0013 \\
  2453989.64105 &          1.0002 &          0.0013 \\
  2453989.64150 &          0.9994 &          0.0013 
\enddata 

\tablecomments{The time stamps represent the Heliocentric Julian Date
  at the time of mid-exposure. The uncertainty estimates are based on
  the procedures described in \S~2. We intend for this Table to appear
  in entirety in the electronic version of the journal. A portion is
  shown here to illustrate its format. The data are also available in
  digital form from the authors upon request.}

\end{deluxetable}

\begin{deluxetable}{lcc}

\tabletypesize{\small}
\tablecaption{System Parameters of TrES-2\label{tbl:params}}
\tablewidth{4.5in}

\tablehead{
\colhead{Parameter} & \colhead{Value} & \colhead{Uncertainty}
}

\startdata
            $(R_\star/R_\odot)(M_\star/0.98~M_\odot)^{-1/3}$& $          1.003$ & $          0.017 $\\
            $(R_p/R_{\rm Jup})(M_\star/0.98~M_\odot)^{-1/3}$& $          1.222$ & $          0.028 $\\
                                           $R_\star/R_\odot$& $          1.003$ & $          0.027 $\\
                                           $R_p/R_{\rm Jup}$& $          1.222$ & $          0.038 $\\
                                             $R_p / R_\star$& $         0.1253$ & $         0.0010 $\\
                                                 $(R_p/a)^2$& $       0.000270$ & $       0.000012 $\\
                                           $M_p/M_{\rm Jup}$\tablenotemark{1}& $          1.198$ & $          0.053 $\\
                                           $T_{\rm eff}$~[K]\tablenotemark{1}& $          5850  $  & $        50 $\\
                                                 $a/R_\star$& $           7.63$ & $           0.12 $\\
                                                   $i$~[deg]& $          83.57$ & $           0.14 $\\
                                                         $b$& $         0.8540$ & $         0.0062 $\\

                               $t_{\rm IV} - t_{\rm I}$~[hr]& $          1.840$ & $          0.020 $\\
                               $t_{\rm II} - t_{\rm I}$~[hr]& $          0.683$ & $          0.045 $\\
                                              $T_c(13)$~[HJD]& $  2453989.75286$ & $        0.00029 $\\
                                              $T_c(15)$~[HJD]& $  2453994.69393$ & $        0.00031 $\\
                                             $T_c(34)$~[HJD]& $  2454041.63579$ & $        0.00030 $
\enddata

\tablenotetext{1}{Adopted from Sozzetti et al. (2007).}

\tablecomments{The system parameters and their associated
  uncertainties for TrES-2 are listed.  $t_{\rm I}$, $t_{\rm II}$, and
  $t_{\rm IV}$, correspond to the times of the first, second, and fourth
  points when the projected limb of the planet contacts that of the
  star.}

\end{deluxetable}

\end{document}